\documentclass[prd,nopacs,preprintnumbers,twocolumn,amsmath,nofootinbib,amssymb]{revtex4}
\usepackage{graphicx,color,dcolumn,booktabs,bm,makecell}
\usepackage{longtable,lscape}
\usepackage{txfonts}
\usepackage{overpic}
\usepackage{epstopdf}
\usepackage{indentfirst}
\usepackage{feynmf}   
\usepackage{slashed}  
\usepackage{cases}
\usepackage{color}
\usepackage{soul}
\usepackage{cancel}
\usepackage{float}
\usepackage{makecell}
\usepackage{gensymb}
\usepackage{multirow}
\usepackage{tikz}
\usepackage{epsfig,dsfont,amssymb,amsmath,amsfonts,amsbsy,mathrsfs}

\graphicspath{{Figures/}} %

\usepackage{hyperref}
\hypersetup{colorlinks,citecolor=blue,anchorcolor=red,menucolor=red,linkcolor=red,filecolor=red,runcolor=red,urlcolor=blue,frenchlinks=red}

\makeatletter
\@addtoreset{equation}{section}
\makeatother

\allowdisplaybreaks

\newcolumntype{I}{!{\vrule width 0.9pt}}

\begin{document}

\title{Newly observed $a_0(1817)$ as the scaling point of constructing the scalar meson spectroscopy}

\author{Dan Guo$^{1,2}$}\email{guod13@lzu.edu.cn}
\author{Wei Chen$^{4}$}\email{chenwei29@mail.sysu.edu.cn}
\author{Hua-Xing Chen$^5$}\email{hxchen@seu.edu.cn}
\author{Xiang Liu$^{1,2,3}$}\email{xiangliu@lzu.edu.cn}
\author{Shi-Lin Zhu$^6$}\email{zhusl@pku.edu.cn}

\affiliation{$^1$School of Physical Science and Technology, Lanzhou University, Lanzhou 730000, China\\
$^2$Research Center for Hadron and CSR Physics, Lanzhou University and Institute of Modern Physics of CAS, Lanzhou 730000, China\\
$^3$Lanzhou Center for Theoretical Physics, Key Laboratory of Theoretical Physics of Gansu Province and Frontier Science Center for Rare Isotopes, Lanzhou University, Lanzhou 730000, China\\
$^4$School of Physics, Sun Yat-Sen University, Guangzhou 510275, China\\
$^5$School of Physics, Southeast University, Nanjing 210094, China\\
$^6$School of Physics and Center of High Energy Physics, Peking University, Beijing 100871, China}

\begin{abstract}
Stimulated by the newly observed $a_0(1817)$ by the BESIII Collaboration, we find a perfect Regge trajectory composed of the $a_0(980)$, $a_0(1450)$, and $a_0(1817)$, which leads us to  
categorize the $a_0(980)$, $a_0(1450)$, and $a_0(1817)$ into the isovector scalar meson family. This scenario is supported by their two-body Okubo-Zweig-Iizuka allowed strong decay behaviors. In this scheme, we also predict the third radial excitation of the $a_0(980)$, which is denoted as the $a_0(2115)$, accessible at future experiment as a direct test of this assignment. We find another Regge trajectory 
which contains three isoscalar scalar states $f_0(980)$, $X(1812)$, and $f_0(2100)$. We investigate their 
two-body Okubo-Zweig-Iizuka allowed strong decay patterns, which are roughly consistent with the experimental data. The $f_0(980)$, $X(1812)$, and $f_0(2100)$ can be well grouped into the isoscalar scalar meson family. We want to emphasize that these two Regge trajectories have a similar slope. In summary, the present work provides a scheme of constructing the scalar meson family based on these reported scalar states. The possibility of the $f_0(1710)$ as the candidate of the scalar glueball cannot be excluded by the observation of the $a_0(1817)$ since the $a_0(1817)$ is more suitable as the isovector partner of the $X(1812)$.
\end{abstract}

\maketitle

\section{introduction}
\label{sec1}

How to quantitatively depict the nonperturbative behavior of quantum chromodynamics (QCD) is an important issue full of challenges and opportunities in the frontiers of particle physics. The study of the hadron spectroscopy can provide valuable hints to deepen our understanding of nonperturbative QCD.

In the hadron zoo, there exist different hadron configurations including the conventional meson and baryon, multiquark state, hybrid and glueball \cite{Gell-Mann:1964ewy,Zweig:1964ruk,Chen:2005mg}.
Due to the observations of the charmoniumlike $XYZ$ states and $P_c$ states, big progress in the search of the multiquark states has been made in the past two decades \cite{Liu:2013waa,Chen:2016qju,Liu:2019zoy,Chen:2022asf}. Early this year, the reported $\eta_1(1855)$
in the $J/\psi\to \gamma\eta\eta^\prime$ decay from the BESIII Collaboration aroused theorists' growing interest in the hybrid meson \cite{BESIII:2022riz,BESIII:2022qzu}.

As a special and key Jigsaw of the hadron family, the glueball
deserves special attention \cite{Klempt:2007cp}.
Usually, the $f_0(1710)$ was suggested as a good candidate of the scalar glueball or at least it has a large glueball component \cite{Cheng:2006hu,Gui:2012gx,Janowski:2014ppa,Ochs:2013gi,Cheng:2015iaa}, which was supported by
the experimental data of $B(f_0(1710)\to \eta\eta^\prime)/B(f_0(1710)\to \pi\pi)=1.61\times 10^{-3}$ from BESIII \cite{BESIII:2022riz,BESIII:2022qzu}. Identifying $f_0(1710)$
as a scalar glueball should be checked from different perspectives. In fact, searching for the isovector partner of the $f_0(1710)$ may provide a criterion to shed light on the nature of $f_0(1710)$. Along this line, in Ref. \cite{BaBar:2021fkz}, the BaBar Collaboration analyzed the $\eta_c\to \eta\pi^+\pi^-$ decay and found a new state $a_0(1710)$ in its $\eta\pi$ decay mode, which has a mass $1704\pm 5(stat.)\pm 2(syst.)$ MeV and width $\Gamma=110\pm 15(stat)\pm 11(stat.)$ MeV \cite{BaBar:2021fkz}. It is natural to assign the $a_0(1710)$ as the partner of $f_0(1710)$, which makes the identification of the $f_0(1710)$ as a scalar glueball ambiguous. Thus, confirming the $a_0(1710)$ by other experiments becomes an urgent issue. 

Very recently, the BESIII Collaboration announced the observation of an isovector scalar state $a_0(1710)$ in the $K_s^0 K^+$ invariant mass spectrum of the $D_s^+\to K_s^0K^+\pi^0$ decay, where the statistical significance of $D_s^+\to a_0(1710)\pi^0$
is larger than $10\sigma$ \cite{BESIII:2022wkv}. The resonance parameters of $a_0(1710)$ are $M=1.817\pm0.008(stat.)\pm0.020(syst.)$ GeV and $\Gamma=0.097\pm0.022(stat.)\pm0.015(syst.)$ GeV \cite{BESIII:2022wkv}. After observing the $a_0(1710)$, the authors of Ref. \cite{Zhu:2022wzk} studied the process $D^+_s \to \pi^+ K^0_S K^0_S$ and the isovector partner of the $f_0(1710)$. 

We notice the obvious difference of the mass values given by BaBar \cite{BaBar:2021fkz} and BESIII \cite{BESIII:2022wkv}. Thus, this mass difference should not be ignord. For this newly observed isovector state \cite{BESIII:2022wkv}, naming it as the $a_0(1817)$ is more appropriate. 

Focusing on this new measurement from BESIII \cite{BESIII:2022wkv}, we should mention the previous observation of the isoscalar state $X(1812)$ in $J/\psi\to \gamma \omega\phi$ by BESII in 2006 \cite{BES:2006vdb}, which has a mass $1812^{+19}_{-26}(stat.)\pm18(syst.)$ MeV and width $105\pm 20(stat.)\pm28(syst.)$ MeV \cite{BES:2006vdb}.
Obviously, the mass of the $a_0(1817)$ given by BESIII \cite{BESIII:2022wkv} is very close to that of the $X(1812)$ \cite{BES:2006vdb}. To some extent, it is more suitable to compare the isovector state $a_0(1817)$ with the isoscalar state $X(1812)$. Thus, the possibility of assigning the $f_0(1710)$ as the scalar glueball cannot be excluded by the observed $a_0(1817)$ signal as claimed by BESIII \cite{BESIII:2022wkv}.

\begin{figure}[htbp]
  \centering
  \begin{tabular}{cc}
\includegraphics[width=0.23\textwidth]{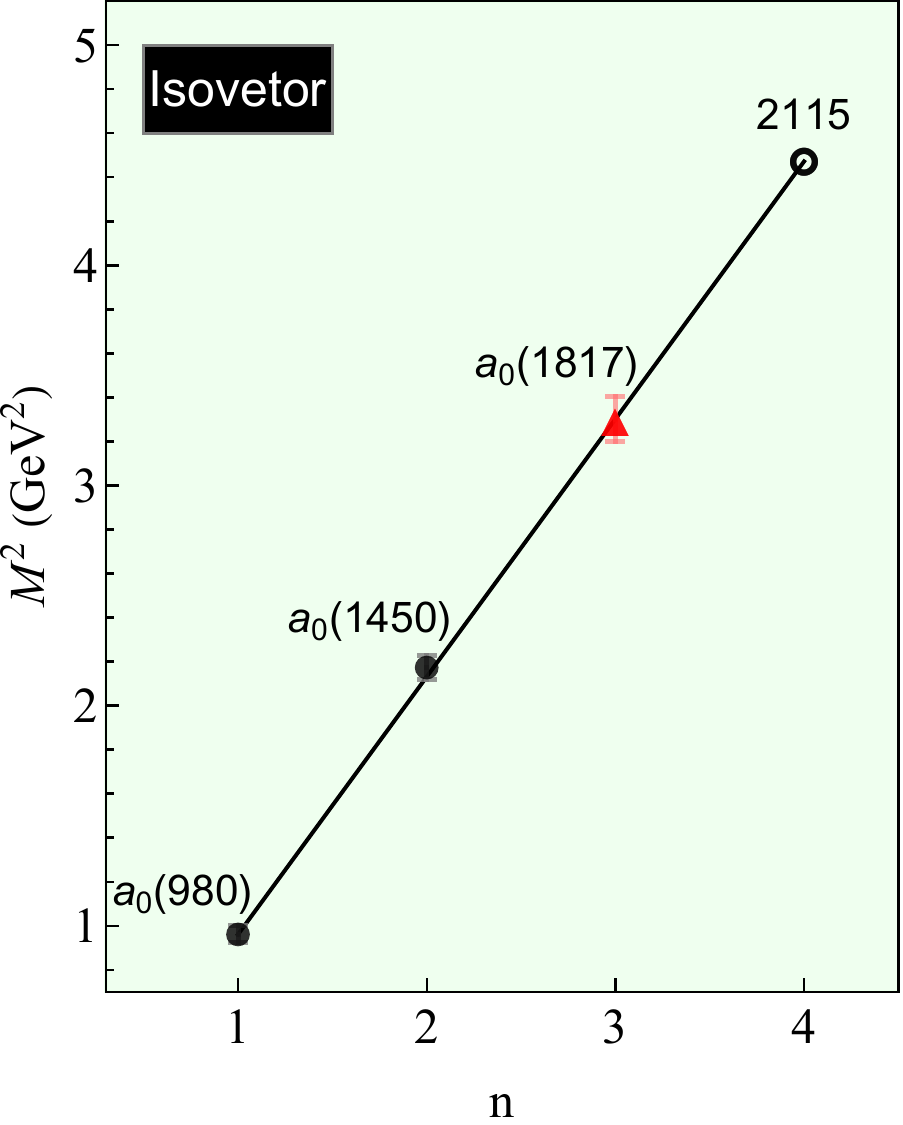}&
\includegraphics[width=0.23\textwidth]{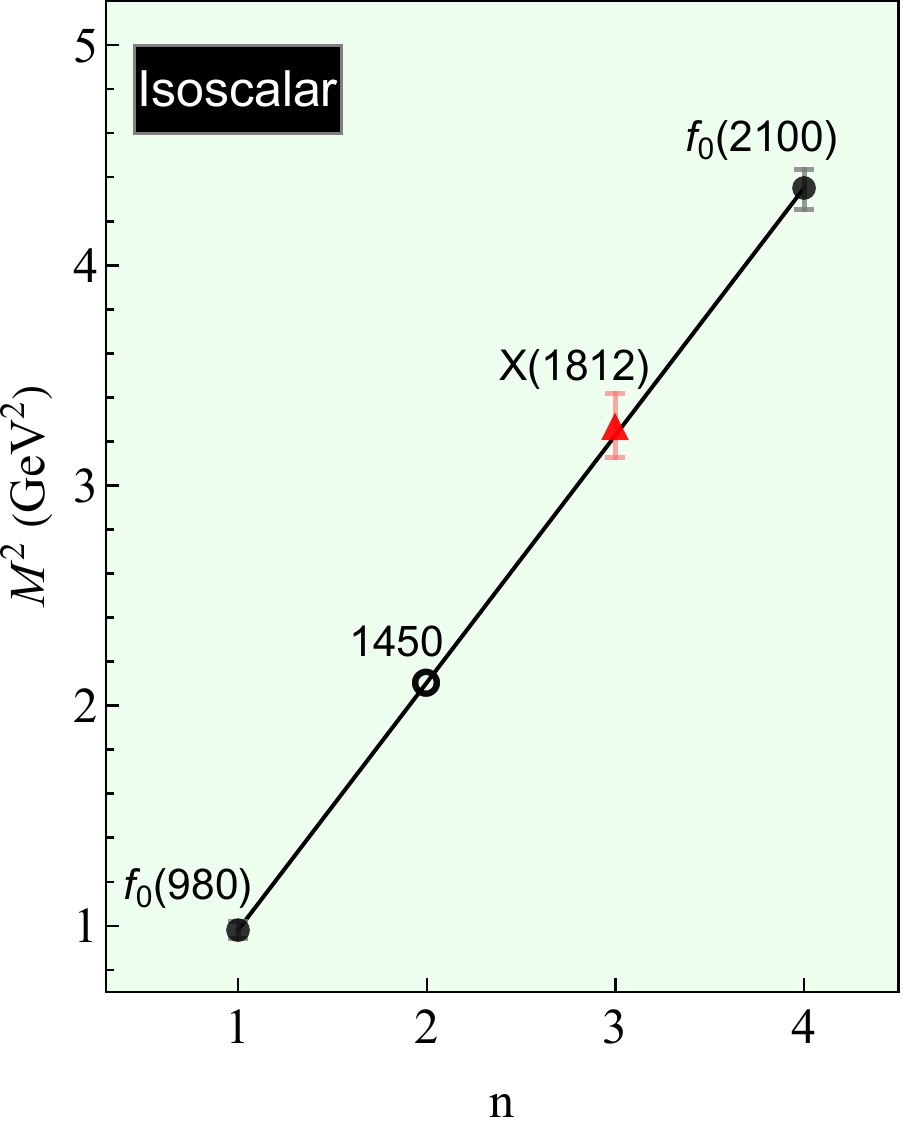}\\

  \end{tabular}
  \caption{Two Regge trajectories for the isovector and isoscalar scalar states. Here, the red triangles denote the $a_0(1817)$ and $X(1812)$, while the solid points and empty circles are the experimental and predicted states, respectively. Except the $a_0(1817)$, $X(1812)$ and predicted $a_0(2115)$, $f_0(1450)$, the masses of the other scalar states are taken from PDG. Here, the established states are marked by the black solid points, while the predicted states are denoted by the black circles. The error bars present total experimental uncertainties.
}
  \label{ReggePlot}
\end{figure}

We find an interesting phenomenon when combining the observed $a_0(1817)$ with
the isovector $a_0$ states listed in
Particle Data Group (PDG) \cite{ParticleDataGroup:2020ssz}. The $a_0(980)$ and $a_0(1450)$ are the established states and the $a_0(1950)$ was omitted from the summary table of PDG. The newly observed $a_0(1817)$ together with $a_0(980)$ and $a_0(1450)$ may form the standard Regge trajectory as shown in Fig. \ref{ReggePlot}. Here, the $a_0(980)$ is the ground state, while the
$a_0(1450)$ and $a_0(1817)$ are the first and the second radial excitations of the $a_0(980)$, respectively. 
With this Regge trajectory, we also predict the third radial excitation of the $a_0(980)$, which has a mass $2115$ MeV. Thus, this predicted isovector scalar state is denoted as the $a_0(2115)$. 
Since the analysis of Regge trajectory can reflect the spectrum behavior of the conventional light flavor mesons \cite{Chew:1962eu,Anisovich:2000kxa}, the observed Regge trajectory may provide direct evidence that these isovector scalar states are suitable to be grouped into the scalar meson family.

For the isoscalar scalar states $f_0(980)$, $X(1812)$, and $f_0(2100)$\footnote{In PDG \cite{ParticleDataGroup:2020ssz}, two $f_0$ states ($f_0(2020)$ and $f_0(2200)$) are omitted from the summary table. In our Regge trajectory analysis, we do not include them.}, we find another Regge trajectory which is also presented in Fig. \ref{ReggePlot}, from which we may give the mass value (1450 MeV) of the first radial excitation of the $f_0(980)$. Obviously, there are two candidates of the first radial excitation of the $f_0(980)$, which are $f_0(1370)$ and $f_0(1500)$. In the following, we will present their decay behavior and discuss which one is more favorable as the first radial excitation of $f_0(980)$.

We can depict the obtained Regge trajectories with the relation
$M^2=M_0^2+(n-1)\mu^2 $, where $M_0$ is the mass of the ground state and $M$ denotes the mass of excited state with the radial quantum number $n$. The slope of the trajectory $\mu^2$ is determined to be $1.17$ GeV$^2$ and $1.12$ GeV$^2$ corresponding to the isovector and isoscalar scalar Regge trajectories, respectively, which show the similarity of these two Regge trajectories. We should emphasize that the newly observed $a_0(1817)$ plays the role of the scaling point when constructing the scalar meson family.

\begin{figure*}
  \centering
  \input{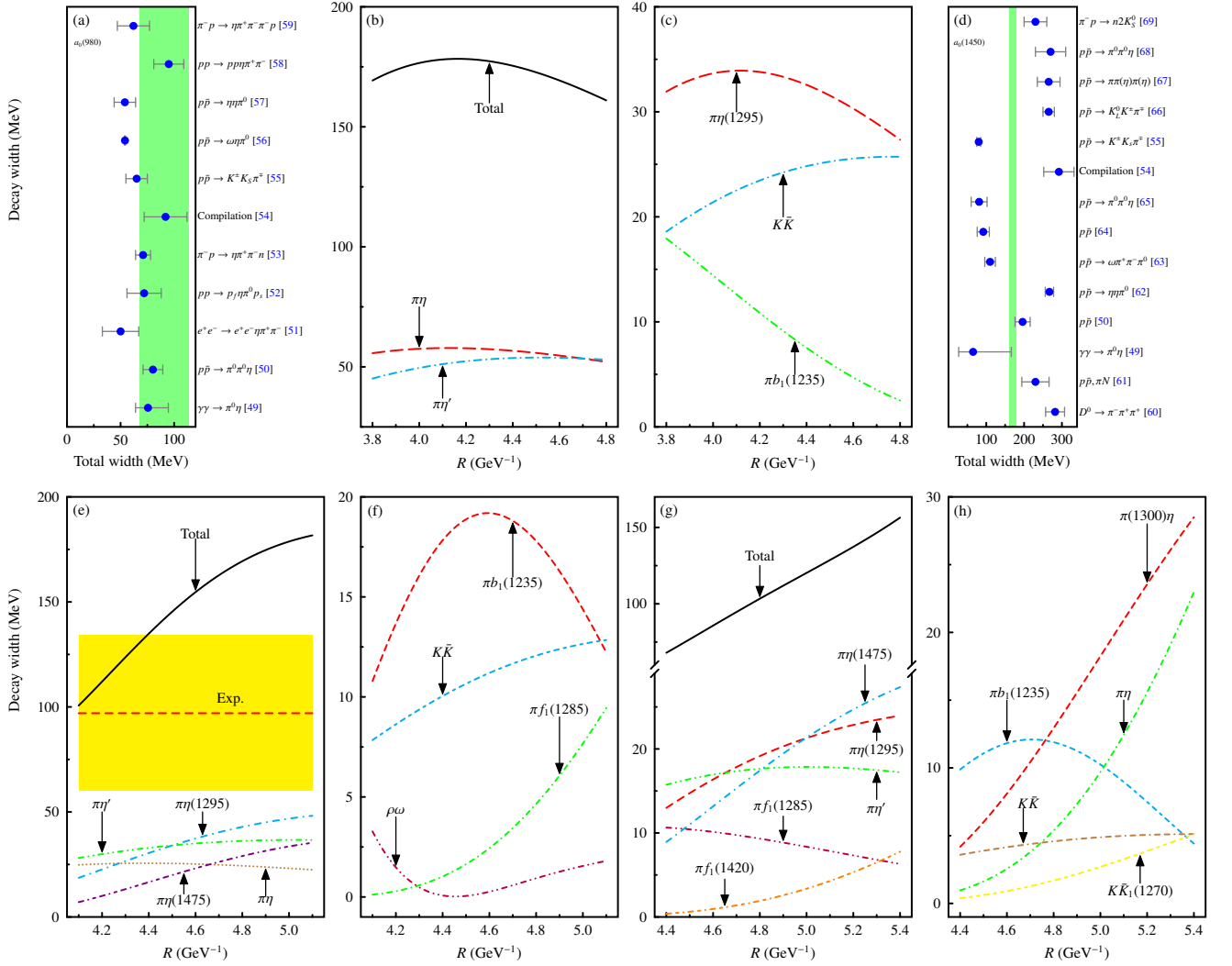}
  \caption{The partial widths and total decay widths of the $a_0(980)$ (a), $a_0(1450)$ (b-c), $a_0(1817)$ (e-f) and the predicted $a_0(2115)$ (g-h) dependent on $R$ values. Here, we compare our results with the experimental width for some states. Especially, for the $a_0(980)$ and $a_0(1450)$, we list different experimental data and the comparison with the calculated total decay width depicted with the green band as shown in diagrams (a) and (d), respectively. In diagram (e), the yellow band denotes the measured width of the $a_0(1817)$ \cite{BESIII:2022wkv}. The channels with widths less than 5 MeV are omitted in picture, but contributions are included in total widths.
}
  \label{a0width}
\end{figure*}

\begin{figure*}
  \centering
  \input{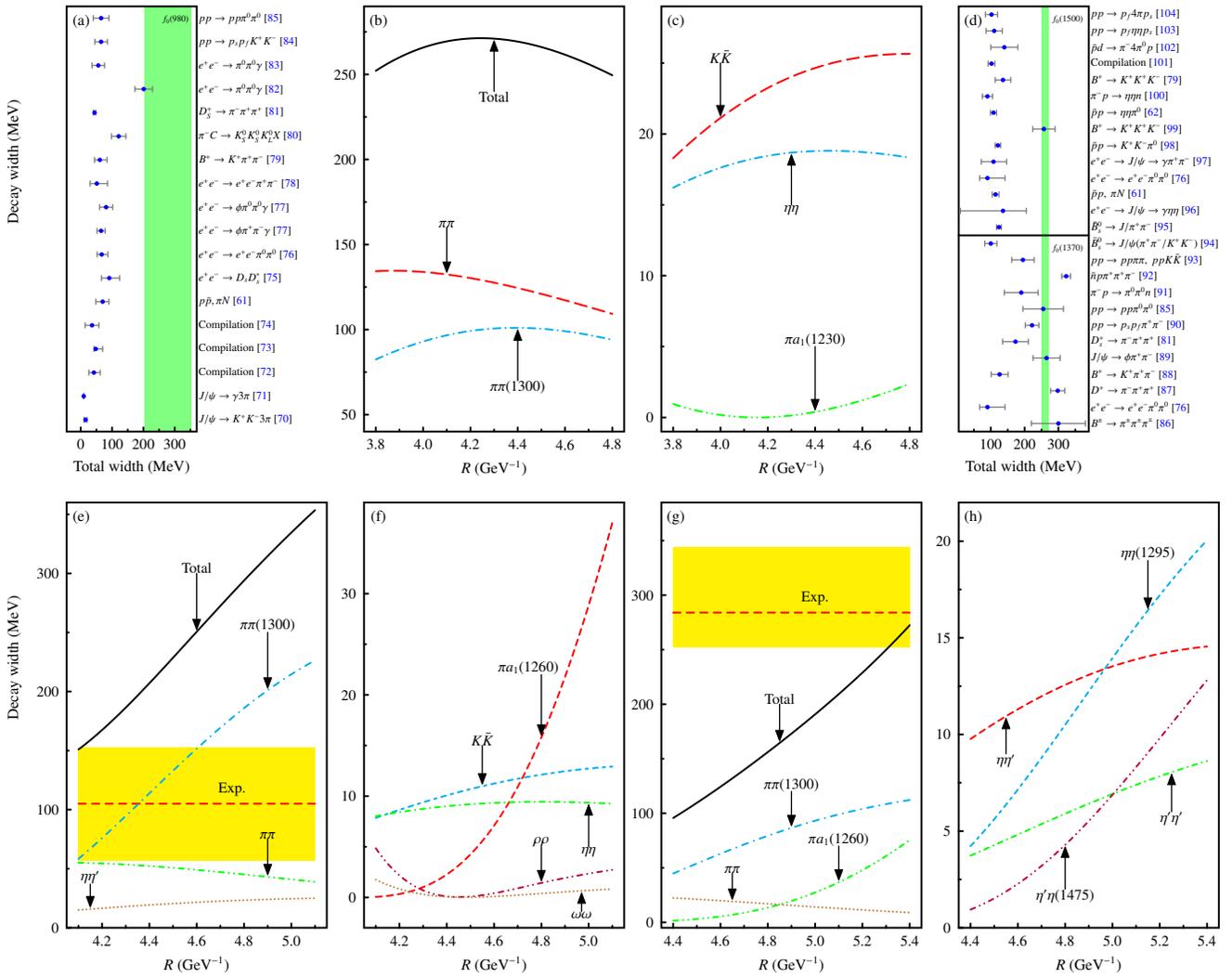}
  \caption{The partial widths and total decay widths for the $f_0(980)$ (a), the first radial excitation of $f_0(980)$ (b-c), $X(1812)$ (e-f) and $f_0(2100)$ (g-h). To show the difference of the experimental width of the $f_0(980)$, we collect them together for comparison with our theoretical result depicted with green band (see digaram (a)). 
  When discussing whether the $f_0(1370)$ and $f_0(1500)$ can be the candidate of the first radial excitation of the $f_0(980)$, we adopt similar treatment as shown in diagram (d). Additionally, the yellow bands denote the measured width of the $X(1812)$ \cite{BES:2006vdb} and $f_0(2100)$ \cite{ParticleDataGroup:2020ssz}. The channels with widths less than 5 MeV are omitted in picture, but contributions are included in total widths.
}
  \label{f0width}
\end{figure*}

\section{Strong decay behaviors}

In order to further examine the above assignment of these discussed scalar states, we study their two-body Okubo-Zweig-Iizuka (OZI) allowed strong decay behaviors. When calculating these partial decay widths, we adopt the quark pair creation (QPC) model \cite{Micu:1968mk,LeYaouanc:1972vsx,LeYaouanc:1973ldf,LeYaouanc:1974cvx,LeYaouanc:1977fsz,LeYaouanc:1977gm} which is extensively applied to study the strong decay of the hadrons \cite{Sun:2009tg,Sun:2010pg,Yu:2011ta,Wang:2012wa,He:2013ttg,Ye:2012gu,Wang:2014sea,Pang:2014laa,Pang:2015eha,Song:2015nia,Song:2015fha,Chen:2015iqa,Pang:2017dlw,Pang:2018gcn,Guo:2019wpx,Wang:2019qyy,Wang:2020due}. 

We first introduce the QPC model. For a realistic decay process of the hadron, the  quark-antiquark pair with the quantum number $J^{PC}=0^{++}$ is created from the vacuum. And then, the created quark and antiquark combine with the corresponding antiquark and quark in initial state to form the final states. The transition operator reads \cite{LeYaouanc:1974cvx}
\begin{eqnarray}
\mathcal{T}&=&-3\gamma\sum_m\langle 1m;1-m|00\rangle
\int d{\mathbf k}_3d{\mathbf k}_4\delta^3({\mathbf k}_3+{\mathbf k}_4)
\nonumber\\&&\times\mathcal{Y}_{1m}\left(\frac{{\mathbf k}_3-{\mathbf k}_4}{2}\right)\chi_{1,-m}^{34}\phi_0^{34}\omega_0^{34}d_{3i}^\dag(\mathbf{k}_3)b_{4j}^\dag(\mathbf{k}_4) .
\end{eqnarray}
Here, the creation probability of the quark-antiquark pair from the vacuum is represented by a parameter $\gamma$, which equals to 7.1 for the $u\bar{u}$ or $d\bar{d}$ quark pair, and $7.1/\sqrt{3}$ for the $s\bar{s}$ quark pair creation from our previous work \cite{Guo:2019wpx}. 
$\phi_0^{34}=(u\bar{u}+d\bar{d}+s\bar{s})/\sqrt{3}$,  $\omega_0^{34}=\delta_{\alpha_3\alpha_4}/\sqrt{3}$ ($\alpha=1,2,3$), and $\chi_{1,-m}^{34}$ are the flavor singlet, the color singlet, and the spin triplet state wave functions, respectively. $i$ and $j$ denote the $SU(3)$ color indices of the created quark pairs from the vacuum. Additionally, we have the definition of the $\ell$th solid harmonic polynomial, i.e., $\mathcal{Y}_{\ell m}(\mathbf{k})=|\mathbf{k}|^\ell Y_{\ell m}(\theta,\phi)$. 
The corresponding transition matrix in the C.M. frame of the $A\to BC$ process is $\langle BC | \mathcal{T}| A\rangle = \delta^3(\mathbf{P}_B+\mathbf{P}_C) \mathcal{M}^{M_{J_A}M_{J_B}M_{J_C}}$ in terms of the helicity amplitude.

The decay amplitude is written in terms of the partial-wave amplitude $\mathcal{M}_{LS}(\mathbf{P})$ \cite{Jacob:1959at},
\begin{eqnarray}
\mathcal{M}_{LS}(\mathbf{P})&=&\frac{4\pi\sqrt{2L+1}}{2J_A+1}\sum_{M_{J_B},M_{J_C}}\langle L0SM_{J_A}|J_A M_{J_A}\rangle\nonumber\\&&\times
\langle J_B M_{J_B}J_C M_{J_C}|SM_{J_A}\rangle \mathcal{M}^{M_{J_A}M_{J_B}M_{J_C}}.
\end{eqnarray}
With the above preparation, the partial decay width reads 
\begin{eqnarray}
\Gamma_i=\pi^2\frac{|\mathbf{P}|\mathcal{S}}{M_A^2}
\sum_{LS}|\mathcal{M}_{LS}|^2,
\end{eqnarray}
where $\mathcal{S}$ is the statistic factor \cite{Guo:2019wpx}.
In the calculation, the harmonic oscillator (HO) wave function is introduced to depict the spatial wave function of these discussed states, which has the general expression $\psi_{n\ell m}(R,\mathbf{p})=\mathcal{R}_{n\ell}(R,\mathbf{p})Y_{\ell m}(\mathbf{p})$. 
Here, the $R$ parameter of these involved mesons can be determined by reproducing the root-mean-square radius of the meson states \cite{Close:2005se}. 
In discussing the dependence of the decay behavior of these scalar states on $R$ value, we take the corresponding $R$ ranges which are determined by 
giving central values and considering variation of $\pm 0.5$ GeV$^{-1}$.

\subsection{Isovector scalar mesons}

As the ground state of the isovector scalar meson family, the $a_0(980)$ can decay into $\pi\eta$, which has the main contribution\footnote{In the present work, we only take the central mass value of the $a_0(980)$ to present its OZI-allowed strong decay behavior. In fact, the $a_0(980)$ may decay into $K\bar{K}$ and $\eta^\prime
\pi$ if considering the width effect of the $a_0(980)$, which are not discussed in the present study.} to the total decay width of the $a_0(980)$. From Fig. \ref{a0width} (a), we show the calculated total width of the $a_0(980)$ dependent on $R$ value. In PDG \cite{ParticleDataGroup:2020ssz}, the average value of the width of the $a_0(980)$ is $50\sim 100$ MeV. Our calculated result roughly overlaps with this experimental range, which can be as a basic test. Thus, it seems reasonable to put the $a_0(980)$ as the ground state. 

For the $a_0(1450)$, we obtain its partial width and total decay width dependent on the $R$ value of the $a_0(1450)$. The main decay modes include the $\pi\eta$, $\pi\eta^\prime$, $\pi\eta(1295)$. In addition, the $K\bar{K}$ also has a sizable contribution to the total decay width. 
The calculated branching ratios of $a_0(1450)\to \pi\eta$, $\pi\eta^\prime$, and $K\bar{K}$ are $(32.3\sim 32.8)\%$, $(26.6\sim 32.9)\%$, and $(11.0\sim 16.0)\%$ corresponding to the $R=3.8\sim4.8$ GeV$^{-1}$ range, which are roughly comparable with the experimental data $\mathcal{B}(a_0(1450)\to \pi\eta)=(9.3\pm2.0)\%$, $\mathcal{B}(a_0(1450)\to \pi\eta^{\prime})=(3.3\pm1.7)\%$, and $\mathcal{B}(a_0(1450)\to K\bar{K})=(8.2\pm2.8)\%$ \cite{ParticleDataGroup:2020ssz}, respectively. 
The obtained total decay width of the $a_0(1450)$ is $161\sim178$ MeV, which is not sensitive to the $R$ value. We notice that the experimental data of the width of $a_0(1450)$ are different from different experimental measurements, which reflects that the width of $a_0(1450)$ is not determined well in experiment. 
Thus, we make a comparison of our theoretical result and these experimental data \cite{ParticleDataGroup:2020ssz}, which shows the consistency of our result with some experimental data \cite{Belle:2009xpa,Bugg:2008ig}.
More precise measurement of the resonance parameters of the $a_0(1450)$ should be paid more attention in future experiment. 

In the following, we focus on the newly observed $a_0(1817)$ and list the partial decay widths and total decay width in Fig. \ref{a0width} (e)-(f). Indeed, the $K\bar{K}$ has a sizable contribution to the total decay width of the $a_0(1817)$, which can explain why the $a_0(1817)$ state can be found in the $K_s^0 K^+$ invariant mass spectrum of the $D_s^+\to K_s^0K^+\pi^0$ decay \cite{BESIII:2022wkv}. 
At present, the decay information of the $a_0(1817)$ is still scarce.
We should emphasize that the experimental total width of the $a_0(1817)$ can be reproduced in the given $R$ range. Thus, the $a_0(1817)$ is a good candidate of the second radial excitation of the $a_0(980)$. The predicted OZI-allowed decay behavior of $a_0(1817)$ is valuable to future experimental study of the $a_0(1817)$. The main decay modes of the $a_0(1817)$ include $\pi\eta(1295)$, $\pi\eta'$, $\pi\eta$, $\pi\eta(1475)$ and $\pi b_1(1235)$. 

In this work, we also predict the third radial excitation of the $a_0(980)$, which is denoted as the $a_0(2115)$. The two-body OZI-allowed decay behavior of the $a_0(2115)$ is given in Fig. \ref{a0width} (g)-(h). The obtained total decay width of the $a_0(2115)$ is $68\sim156$ MeV corresponding to $R=4.4\sim5.4$ GeV$^{-1}$. Among these allowed decays, we suggest three typical channels $\pi\eta$, $\pi\eta^\prime$, and $K\bar{K}$ with sizable partial widths in the experimental search for the $a_0(2115)$. If the $a_0(2115)$ can be observed in the future, it will be a good test of the construction of isovector scalar meson family proposed in this work.  

\subsection{Isoscalar scalar mesons}

With the perfect Regge trajectory for the isovector mesons with the $a_0(980)$, $a_0(1450)$ and $a_0(1817)$, we study their two-body OZI-allowed strong decays, which further support 
the scenario of categorizing them into the isovector meson family. The isovector meosns are always accompanied by their isoscalar partners. So there must exist a similar Regge trajectory for the isoscalar mesons. As discussed in Sec. \ref{sec1}, a corresponding Regge trajectory for isoscalar mesons with the $f_0(980)$, $X(1812)$, and $f_0(2100)$ does exist. Under this assignment, we will discuss their two-body OZI-allowed strong decays.

Treating the $f_0(980)$ as the ground state of the isoscalar scalar meson family, the $f_0(980)$ decays into $\pi\pi$\footnote{Similar to the case of the $a_0(980)$, we do not consider the $K\bar{K}$ mode.}. And its total width is $203\sim351$ MeV corresponding to $R=3.3\sim4.3$ GeV$^{-1}$ from the QPC model. For the $f_0(980)$, its width is not well determined experimentally  \cite{ParticleDataGroup:2020ssz}. Thus, we  compare the calculated total decay width with these concrete experimental values of the $f_0(980)$ width as shown in Fig. \ref{f0width} (a). Generally, our obtained total decay width of the $f_0(980)$ roughly overlaps with some experimental data \cite{Achasov:2000ym}. 

We should mention the decay behavior of the $X(1812)$. The obtained total decay width of the $X(1812)$ is dependent on $R$ value as shown in Fig. \ref{f0width} (e)-(f). Here, the $\pi\pi(1300)$ and $\pi\pi$ are two main decay channels, while the $\eta\eta'$, $\pi a_1(1260)$ and $K\bar{K}$ have sizable contributions to the total decay width of the $X(1812)$ with
$R=4.1\sim5.1$ GeV$^{-1}$. Since the $X(1812)$ was observed in its OZI-forbidden decay channel $\omega\phi$ \cite{BES:2006vdb}, we are waiting for further experimental observation of its OZI-allowed decay channels. Thus, these obtained decay information may provide valuable information in the experimental search for the $X(1812)$. According to the present experimental information, we may conclude that it is suitable to assign the
$X(1812)$ as the second radial excitation of the $f_0(980)$. 

In this work, we also illustrate the decay behavior of the $f_0(2100)$, which is a scalar state omitted from the summary table of PDG \cite{ParticleDataGroup:2020ssz}. The calculated total decay width can reach the experimental width of the $f_0(2100)$ when taking the allowed $R=4.4\sim5.4$ GeV$^{-1}$ range. 
The $\pi\pi(1300)$, $\pi a_1(1260)$ and $\pi\pi$ are its main decay channels, and its sizable decays are collected in Fig. \ref{f0width} (h). Experimentally establishing the $f_0(2100)$ will be an interesting task.

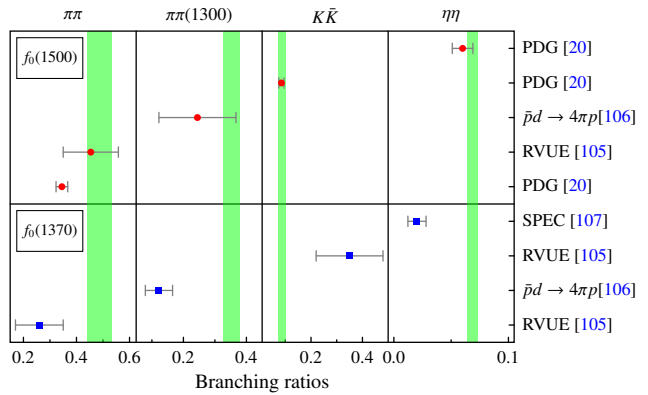
\begin{figure}
\centering
\begin{tikzpicture}
\node[below] at (3.85,0.189455) {\scalebox{0.85}{Branching ratios}};
\node[] at (1.3375,4.84763) {\scalebox{0.75}{$\pi\pi$}};
\node[] at (3.0125,4.84763) {\scalebox{0.75}{$\pi\pi(1300)$}};
\node[] at (4.6875,4.84763) {\scalebox{0.75}{$K\bar{K}$}};
\node[] at (6.3625,4.84763) {\scalebox{0.75}{$\eta\eta$}};
\draw[line width=0.50pt,line cap=round] (0.5,2.34027)--(7.2,2.34027);
\draw[line width=0.50pt,line cap=round] (2.175,0.5)--(2.175,4.6406);
\draw[line width=0.50pt,line cap=round] (3.85,0.5)--(3.85,4.6406);
\draw[line width=0.50pt,line cap=round] (5.525,0.5)--(5.525,4.6406);
\draw[] (0.58777,4.55283)--(1.46547,4.55283)--(1.46547,4.01041)--(0.58777,4.01041)--cycle;
\node[] at (1.02662,4.28162) {\scalebox{0.67}{$f_0(1500)$}};
\draw[] (0.58777,2.2525)--(1.46547,2.2525)--(1.46547,1.71008)--(0.58777,1.71008)--cycle;
\node[] at (1.02662,1.98129) {\scalebox{0.67}{$f_0(1370)$}};
\draw[line width=0.50pt,line cap=round] (0.676316,0.5)--(0.676316,0.567);
\node[below] at (0.676316,0.5) {\scalebox{0.75}{0.2}};
\draw[line width=0.50pt,line cap=round] (1.38158,0.5)--(1.38158,0.567);
\node[below] at (1.38158,0.5) {\scalebox{0.75}{0.4}};
\draw[line width=0.50pt,line cap=round] (2.08684,0.5)--(2.08684,0.567);
\node[below] at (2.08684,0.5) {\scalebox{0.75}{0.6}};
\draw[line width=0.50pt,line cap=round] (1.02895,0.5)--(1.02895,0.5335);
\draw[line width=0.50pt,line cap=round] (1.73421,0.5)--(1.73421,0.5335);
\draw[line width=0.50pt,line cap=round] (2.80313,0.5)--(2.80313,0.567);
\node[below] at (2.80313,0.5) {\scalebox{0.75}{0.2}};
\draw[line width=0.50pt,line cap=round] (3.64063,0.5)--(3.64063,0.567);
\node[below] at (3.64063,0.5) {\scalebox{0.75}{0.4}};
\draw[line width=0.50pt,line cap=round] (2.38438,0.5)--(2.38438,0.5335);
\draw[line width=0.50pt,line cap=round] (3.22188,0.5)--(3.22188,0.5335);
\draw[line width=0.50pt,line cap=round] (4.49949,0.5)--(4.49949,0.567);
\node[below] at (4.49949,0.5) {\scalebox{0.75}{0.2}};
\draw[line width=0.50pt,line cap=round] (5.18316,0.5)--(5.18316,0.567);
\node[below] at (5.18316,0.5) {\scalebox{0.75}{0.4}};
\draw[line width=0.50pt,line cap=round] (4.15765,0.5)--(4.15765,0.5335);
\draw[line width=0.50pt,line cap=round] (4.84133,0.5)--(4.84133,0.5335);
\draw[line width=0.50pt,line cap=round] (5.60114,0.5)--(5.60114,0.567);
\node[below] at (5.60114,0.5) {\scalebox{0.75}{0.0}};
\draw[line width=0.50pt,line cap=round] (7.12386,0.5)--(7.12386,0.567);
\node[below] at (7.12386,0.5) {\scalebox{0.75}{0.1}};
\draw[line width=0.50pt,line cap=round] (6.3625,0.5)--(6.3625,0.5335);
\fill[green,opacity=0.5] (1.51558,0.5)--(1.84705,0.5)--(1.84705,4.6406)--(1.51558,4.6406)--cycle;
\fill[green,opacity=0.5] (3.33075,0.5)--(3.55269,0.5)--(3.55269,4.6406)--(3.33075,4.6406)--cycle;
\fill[green,opacity=0.5] (4.06536,0.5)--(4.16791,0.5)--(4.16791,4.6406)--(4.06536,4.6406)--cycle;
\fill[green,opacity=0.5] (6.57568,0.5)--(6.71273,0.5)--(6.71273,4.6406)--(6.57568,4.6406)--cycle;
\draw[line width=0.50pt,line cap=round] (7.2,4.41057)--(7.133,4.41057);
\node[right] at (7.2,4.41057) {\scalebox{0.75}{PDG~\cite{ParticleDataGroup:2020ssz}}};
\draw[line width=0.50pt,line cap=round,gray] (6.37773,4.41057)--(6.65182,4.41057);
\draw[line width=0.50pt,line cap=round,gray] (6.37773,4.34156)--(6.37773,4.47958);
\draw[line width=0.50pt,line cap=round,gray] (6.65182,4.34156)--(6.65182,4.47958);
\fill[red] (6.51477,4.41057)circle(0.0460067);
\draw[line width=0.50pt,line cap=round] (7.2,3.9505)--(7.133,3.9505);
\node[right] at (7.2,3.9505) {\scalebox{0.75}{PDG~\cite{ParticleDataGroup:2020ssz}}};
\draw[line width=0.50pt,line cap=round,gray] (4.07219,3.9505)--(4.14056,3.9505);
\draw[line width=0.50pt,line cap=round,gray] (4.07219,3.88149)--(4.07219,4.01951);
\draw[line width=0.50pt,line cap=round,gray] (4.14056,3.88149)--(4.14056,4.01951);
\fill[red] (4.10638,3.9505)circle(0.0460067);
\draw[line width=0.50pt,line cap=round] (7.2,3.49043)--(7.133,3.49043);
\node[right] at (7.2,3.49043) {\scalebox{0.75}{$\bar{p}d\to 4\pi p$\cite{CRYSTALBARREL:2001ldq}}};
\draw[line width=0.50pt,line cap=round,gray] (2.47755,3.49043)--(3.50139,3.49043);
\draw[line width=0.50pt,line cap=round,gray] (2.47755,3.42142)--(2.47755,3.55944);
\draw[line width=0.50pt,line cap=round,gray] (3.50139,3.42142)--(3.50139,3.55944);
\fill[red] (2.98947,3.49043)circle(0.0460067);
\draw[line width=0.50pt,line cap=round] (7.2,3.03037)--(7.133,3.03037);
\node[right] at (7.2,3.03037) {\scalebox{0.75}{RVUE~\cite{Bugg:1996ki}}};
\draw[line width=0.50pt,line cap=round,gray] (1.20526,3.03037)--(1.93874,3.03037);
\draw[line width=0.50pt,line cap=round,gray] (1.20526,2.96136)--(1.20526,3.09938);
\draw[line width=0.50pt,line cap=round,gray] (1.93874,2.96136)--(1.93874,3.09938);
\fill[red] (1.572,3.03037)circle(0.0460067);
\draw[line width=0.50pt,line cap=round] (7.2,2.5703)--(7.133,2.5703);
\node[right] at (7.2,2.5703) {\scalebox{0.75}{PDG~\cite{ParticleDataGroup:2020ssz}}};
\draw[line width=0.50pt,line cap=round,gray] (1.11005,2.5703)--(1.26521,2.5703);
\draw[line width=0.50pt,line cap=round,gray] (1.11005,2.50129)--(1.11005,2.63931);
\draw[line width=0.50pt,line cap=round,gray] (1.26521,2.50129)--(1.26521,2.63931);
\fill[red] (1.18763,2.5703)circle(0.0460067);
\draw[line width=0.50pt,line cap=round] (7.2,2.11023)--(7.133,2.11023);
\node[right] at (7.2,2.11023) {\scalebox{0.75}{SPEC~\cite{Anisovich:2001ay}}};
\draw[line width=0.50pt,line cap=round,gray] (5.78752,2.11023)--(6.02872,2.11023);
\draw[line width=0.50pt,line cap=round,gray] (5.78752,2.04122)--(5.78752,2.17924);
\draw[line width=0.50pt,line cap=round,gray] (6.02872,2.04122)--(6.02872,2.17924);
\fill[blue] (5.86211,2.06423)rectangle(5.95412,2.15624);
\draw[line width=0.50pt,line cap=round] (7.2,1.65017)--(7.133,1.65017);
\node[right] at (7.2,1.65017) {\scalebox{0.75}{RVUE~\cite{Bugg:1996ki}}};
\draw[line width=0.50pt,line cap=round,gray] (4.56786,1.65017)--(5.45663,1.65017);
\draw[line width=0.50pt,line cap=round,gray] (4.56786,1.58116)--(4.56786,1.71918);
\draw[line width=0.50pt,line cap=round,gray] (5.45663,1.58116)--(5.45663,1.71918);
\fill[blue] (4.96624,1.60416)rectangle(5.05825,1.69617);
\draw[line width=0.50pt,line cap=round] (7.2,1.1901)--(7.133,1.1901);
\node[right] at (7.2,1.1901) {\scalebox{0.75}{$\bar{p}d\to 4\pi p$\cite{CRYSTALBARREL:2001ldq}}};
\draw[line width=0.50pt,line cap=round,gray] (2.29728,1.1901)--(2.65908,1.1901);
\draw[line width=0.50pt,line cap=round,gray] (2.29728,1.12109)--(2.29728,1.25911);
\draw[line width=0.50pt,line cap=round,gray] (2.65908,1.12109)--(2.65908,1.25911);
\fill[blue] (2.43217,1.14409)rectangle(2.52418,1.23611);
\draw[line width=0.50pt,line cap=round] (7.2,0.730033)--(7.133,0.730033);
\node[right] at (7.2,0.730033) {\scalebox{0.75}{RVUE~\cite{Bugg:1996ki}}};
\draw[line width=0.50pt,line cap=round,gray] (0.570526,0.730033)--(1.20526,0.730033);
\draw[line width=0.50pt,line cap=round,gray] (0.570526,0.661023)--(0.570526,0.799043);
\draw[line width=0.50pt,line cap=round,gray] (1.20526,0.661023)--(1.20526,0.799043);
\fill[blue] (0.841888,0.684027)rectangle(0.933901,0.77604);
\draw[line width=0.50pt,line cap=round] (0.5,0.5)--(7.2,0.5)--(7.2,4.6406)--(0.5,4.6406)--cycle;
\end{tikzpicture}
\caption{The comparisons of the calculated branching ratios (green bands) of the $\pi\pi$, $\pi\pi(1300)$, $K\bar{K}$, and $\eta\eta$ channels of the first radial excitation of the $f_0(980)$ with the experimental data for the $f_0(1370)$ and $f_0(1500)$.}
\label{branchratio}
\end{figure}

Taking the mass of the first radial excitation of the $f_0(980)$ to be $1450$ MeV as given by the analysis of Regge trajectory, we show its 
decay behavior in Fig. \ref{f0width} (b)-(c). There are two scalar states, $f_0(1370)$ and $f_0(1500)$, around 1450 MeV. In Fig. \ref{f0width} (d), we compare the theoretical width with the experimental widths of the $f_0(1370)$ and $f_0(1500)$. Our theoretical result of the width of the $f_0(1450)$ is $249\sim 271$ MeV in the $R$ range $3.8\sim 4.8$, which is consistent with some experimental values of the $f_0(1370)$ width measured in Refs. \cite{BaBar:2009vfr,BES:2004twe,GAMS:1999jhw} and one measurement of the $f_0(1500)$ width in Ref. \cite{BaBar:2006hyf}. Among the allowed decay channels, the branching ratios of the 
$\pi\pi$, $\pi\pi(1300)$, $K\bar{K}$ and $\eta\eta$ channels are  $(43.8\sim53.2)\%$, $(32.6\sim37.9)\%$, $(7.3\sim10.3)\%$ and $(6.4\sim7.3)\%$, respectively. In Fig. \ref{branchratio}, we also make a comparison of these results with the corresponding branching ratios associated with the $f_0(1370)$ and $f_0(1500)$. 
In short, the possibility of the $f_0(1370)$ and $f_0(1500)$ as the first radial excitation of 
the $f_0(980)$ cannot be excluded. The precise measurement of the resonance parameters and decay behavior of the $f_0(1370)$ and $f_0(1500)$ will be very helpful.

\section{Discussion and conclusion}

Very recently, BESIII reported the observation of an isovector scalar state $a_0(1817)$ by 
analyzing the $D_s^+\to K_S^0K^+\pi^0$ decay \cite{BESIII:2022wkv}, which stimulates our interest in 
constructing the scalar meson family using the $a_0(1817)$ as the scaling point. 
The $a_0(1817)$, $a_0(980)$, and $a_0(1450)$ form a Regge trajectory, which suggests  
the possibility of allotting the $a_0(1817)$, $a_0(980)$, and $a_0(1450)$ into isovector scalar meson family. Their two-body 
OZI-allowed decay behaviors support this assignment. The predicted higher scalar meson, $a_0(2115)$ can be accessible at future experiment as the test of the above scheme.

The existence of the Regge trajectory of the isovector scalar mesons naturally leads to the existence of the similar Regge trajectory of isoscalar scalar mesons. Along this line, we check these observed isoscalar scalar states collected in PDG \cite{ParticleDataGroup:2020ssz} and indeed find another Regge trajectory composed of the $f_0(980)$, $X(1812)$, and $f_0(2100)$, which are the ground state, the second and the third radially excited states of the isoscalar scalar meson family, respectively. Their two-body OZI-allowed decay behaviors also support our observation. With this Regge trajectory, the first radial excitation of the $f_0(980)$ lies around 1450 MeV, close to the $f_0(1370)$ and $f_0(1500)$. The decay behavior of the $f_0(1450)$ is roughly consistent with the experimental data of the $f_0(1370)$ and $f_0(1500)$. There exists the possibility of the $f_0(1370)$ and $f_0(1500)$ as the candidate of the first radial excitation of the $f_0(980)$.  One may wonder whether the $f_0(1370)$ and $f_0(1500)$ signals arise from the same isoscalar scalar state $f_0(1450)$. Very recently, the BESIII Collaboration reported the simultaneous production of the $f_0(1500)$ and $f_2(1525)$ in the decays of $X(2600)$ \cite{BESIII:2022llk} 
\begin{eqnarray}
\nonumber && \mathcal{B}(J/\psi \to \gamma X(2600)) \times
\mathcal{B}(X(2600) \to f_0(1500) \eta^\prime) \nonuber\\&&\times
\mathcal{B}(f_0(1500) \to \pi^+ \pi^-)= (3.39 \pm 0.18 ^{+0.91}_{-0.66} ) \times 10^{-5} \, ,\nonumber
\\
\nonumber && \mathcal{B}(J/\psi \to \gamma X(2600)) \times
\mathcal{B}(X(2600) \to f_2^\prime(1525) \eta^\prime) \nonumber\\
&&\times
\mathcal{B}( f_2^\prime(1525) \to \pi^+ \pi^-)
= (2.43 \pm 0.13 ^{+0.31}_{-1.11}) \times 10^{-5} \, .\nonumber
\end{eqnarray}
The similar production rate and decay process  strongly indicate that the $f_0(1500)$ and $f_2(1525)$ may have the same inner structures.
Since the tensor glueball is expected to lie well above 2 GeV, the $f_2(1525)$ is very probably a conventional tensor meson. In other words, the $f_0(1500)$ is probably a scalar meson dominated by the $q\bar q$ component, which is consistent with the observation from the Regge analysis. 
In short summary, the newly observed $a_0(1817)$ can be a good isovector partner of the $X(1812)$. Thus, the possibility of the $f_0(1710)$ as the scalar glueball cannot be excluded by the observation of the $a_0(1817)$ \cite{BESIII:2022wkv}.

\section*{ACKNOWLEDGMENTS}
D.G. would like to thank S.Q. L. for helpful discussion. This work is supported by the China National Funds for Distinguished Young Scientists under Grant No. 11825503, National Key Research and Development Program of China under Contract No. 2020YFA0406400, the 111 Project under Grant No. B20063, and the National Natural Science Foundation of China under Grants No. 12047501, 11975033 and No. 12070131001，No. 12175318, and Natural Science Foundation of Guangdong Province of China under Grant No. 2022A1515011922.

\end{document}